\documentclass[%
  journal = jpclcd,%
  layout = twocolumn,%
  ]{achemso}
\usepackage[latin1]{inputenc}
\usepackage{graphicx}
\usepackage{amsmath,amssymb}
\usepackage{amsthm}
\usepackage[version=3]{mhchem}
\usepackage{booktabs}
\usepackage{widetable}
\usepackage{setspace}
\usepackage{multirow}
\usepackage{upgreek}
\usepackage{physics}
\usepackage{braket}
\usepackage{cleveref}
\usepackage{longtable}
\usepackage{braket}

\usepackage{threeparttable}
\usepackage[varg]{txfonts}
\usepackage[font=footnotesize,labelfont=bf]{caption}
\usepackage[small]{titlesec}
\setkeys{acs}{articletitle = true}
\setkeys{acs}{maxauthors=15}
\setkeys{acs}{etalmode=truncate}
\usepackage{etoolbox}
\makeatletter
\patchcmd{\ttlh@hang}{\parindent\z@}{\parindent\z@\leavevmode}{}{}
\patchcmd{\ttlh@hang}{\noindent}{}{}{}
\makeatother

\titleformat{\subsubsection}[runin]
{\normalfont\footnotesize\bfseries}{\thesubsubsection}{0.5em}{}
\titlespacing{\subsubsection}{0.0em}{0.0em}{0.5em}

\crefname{equation}{equation}{Equations}
\Crefname{equation}{Equation}{Equations}
\crefrangelabelformat{equation}{(#3#1#4--#5#2#6)}

\crefmultiformat{equation}{equations (#2#1#3}{, #2#1#3)}{#2#1#3}{#2#1#3}
\Crefmultiformat{equation}{Equations (#2#1#3}{, #2#1#3)}{#2#1#3}{#2#1#3}

\usepackage{xcolor}
\definecolor{aaltoOrange}{RGB}{255,121,0}%
\definecolor{aaltoBlue}{RGB}{0,101,189}%

\title{\Large{Accurate absolute and relative core-level binding energies from \boldmath{$GW$}}}

\author {\normalsize{Dorothea Golze}}
\affiliation{\small{Department of Applied Physics, Aalto University, Otakaari 1, FI-02150 Espoo, Finland}}
\email{dorothea.golze@aalto.fi}
\author {\normalsize{Levi Keller}}
\affiliation{\small{Department of Applied Physics, Aalto University, Otakaari 1, FI-02150 Espoo, Finland}}
\author {Patrick Rinke}
\affiliation{\small{Department of Applied Physics, Aalto University, Otakaari 1, FI-02150 Espoo, Finland}}
\let\oldmaketitle\maketitle
\let\maketitle\relax

\DeclareMathOperator{\sgn}{sgn}

\begin{tocentry}
\centering
\includegraphics[width=1.0\columnwidth]{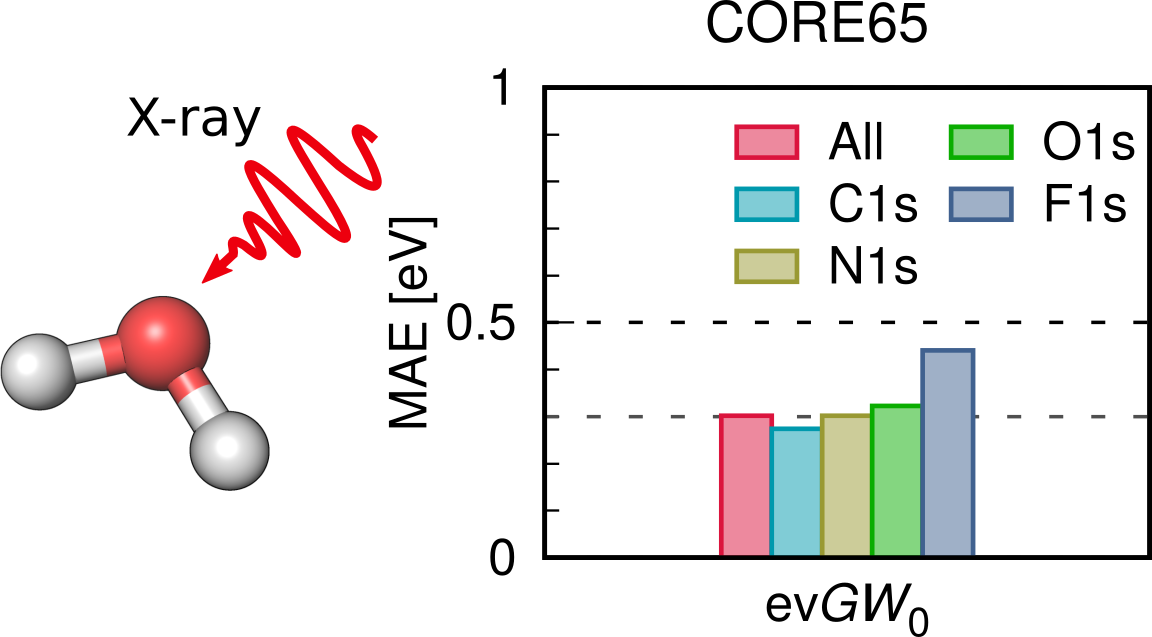}
\end{tocentry}

\begin{document}
\linespread{1.1}
\fontsize{10}{12}\selectfont
\twocolumn[
  \begin{@twocolumnfalse}
    \oldmaketitle
    \begin{abstract}
\fontsize{10}{12}\selectfont  
We present an accurate approach to compute X-ray photoelectron spectra based on the $GW$ Green's function method, that overcomes shortcomings of common density functional theory approaches. $GW$ has become a popular tool to compute valence excitations for a wide range of materials. However, core-level spectroscopy is thus far almost uncharted in $GW$. We show that single-shot perturbation calculations in the $G_0W_0$ approximation, which are routinely used for valence states, cannot be applied for core levels and suffer from an extreme, erroneous transfer of spectral weight to the satellite spectrum. The correct behavior can be restored by partial self-consistent $GW$ schemes or by using hybrid functionals with almost 50\% of exact exchange as starting point for $G_0W_0$. We include also relativistic corrections and present a benchmark study for 65 molecular 1s excitations. Our absolute and relative $GW$ core-level binding energies agree within 0.3 and 0.2 eV with experiment, respectively.
    \end{abstract}
  \end{@twocolumnfalse}
  ]

Core-level spectroscopy techniques, such as X-ray photoelectron spectroscopy (XPS), are important tools for chemical analysis and can be applied to a broad range of systems including 
crystalline\cite{Bagus2013} and amorphous materials~\cite{Sainio2016,Aarva2019a,Aarva2019b}, liquids\cite{Cremer2010,Garcia2011,Santos2015}, adsorbates at surfaces\cite{Fasel2018} or 2D 
materials~\cite{Scardamaglia2017,Susi2018}. XPS measures core-level binding energies (BEs), which are element-specific, but depend on the local chemical environment. For 
complex materials, the assignment of the experimental XPS signals to the specific atomic sites is notoriously difficult, due to overlapping spectral features or the lack of 
well-defined reference data.\cite{Aarva2019a}
Accurate theoretical tools for the prediction of core excitations are therefore important to guide the experiment. Calculated \emph{relative} binding energies, i.e., BE shifts with respect to a reference XPS signal, are particularly useful for the interpretation of experimental spectra. However, the prediction of accurate \emph{absolute} core-level energies is equally important, in particular when reference core-level energies are not available. \par
The most common approach to compute core-level BEs is the Delta self-consistent field ($\Delta$SCF) method, which is based on Kohn-Sham density functional theory (KS-DFT). 
In $\Delta$SCF, the core-level binding energies are calculated as total energy difference between the neutral and the ionized system.\cite{Bagus1965} Relative core-level BEs from 
$\Delta$SCF generally compare well to experiment. For small molecules, deviations typically lie in the range of $0.2-0.3$~eV,\cite{Bellafont2016} which is well within or close to the chemical resolution required for most elements. The dependence of the relative BEs on the exchange-correlation (XC) functional is almost negligible for small systems,\cite{Bellafont2016} 
but can be more severe for complex materials.\cite{Susi2015,Susi2018} Absolute  $\Delta$SCF BEs can differ by several eV from the experimental data. This deviation is quite sensitive to the XC-functional.\cite{Bellafont2015} The best results for absolute core excitations have been obtained using the TPSS\cite{Bellafont2016b} and SCAN\cite{Kahk2019} 
meta-generalized gradient approximations. The reported mean absolute deviations from experiment lie in the range of $\approx0.2$~eV for benchmark sets of small  molecules. For medium-sized to large molecules, however, the accuracy of $\Delta$SCF can quickly reduce by an order of magnitude for absolute BEs.\cite{Golze2018} This behavior can be partly attributed to an insufficient localization of the core hole in the calculation for the ionized system. Constraining the core hole in a particular state can be  difficult and variational instabilities are not uncommon.\cite{Michelitsch2019}

Most importantly, $\Delta$SCF cannot be applied without further approximations to periodic systems, such as surfaces, where the ionized calculation would lead to a Coulomb divergence. \cite{Ozaki2017} Such divergences can be circumvented by using cluster models\cite{Kahk2018}, by neutralizing the unit or supercell with compensating background charges\cite{Bellafont2017} or by adding the compensating electrons to the conduction band.\cite{Pehlke1993,Koehler2004,Olovsson2005} However, these 
approximations can obscure the calculations and even lead to qualitatively wrong results, as recently demonstrated for oxide surfaces\cite{Bagus2019}.\par 
Higher-level theoretical methods such as Delta coupled-cluster ($\Delta$CC) approaches yield highly accurate relative and absolute core ionization energies.\cite{Zheng2019,Holme2011,Sen2018} $\Delta$CC also requires the computation of a core-ionized system leading  to the same conceptual problems as in $\Delta$SCF. Response theories, e.g., equation-of-motion coupled cluster, avoid these problems, but deviate by several eV from experiment and require at least  triples contributions for quantitative agreement.\cite{Liu2019} Good accuracy for deep states was reported for a  recently introduced direct approach based on effective one-particle energies from the generalized KS random phase approximation.\cite{Voora2019} However, 
the application of these higher-level methods is restricted to small or medium-sized systems due to 
unfavorable scaling with system size and large computational prefactors. \par
\begin{figure*}[t]
\centering
  \includegraphics[width=0.99\textwidth]{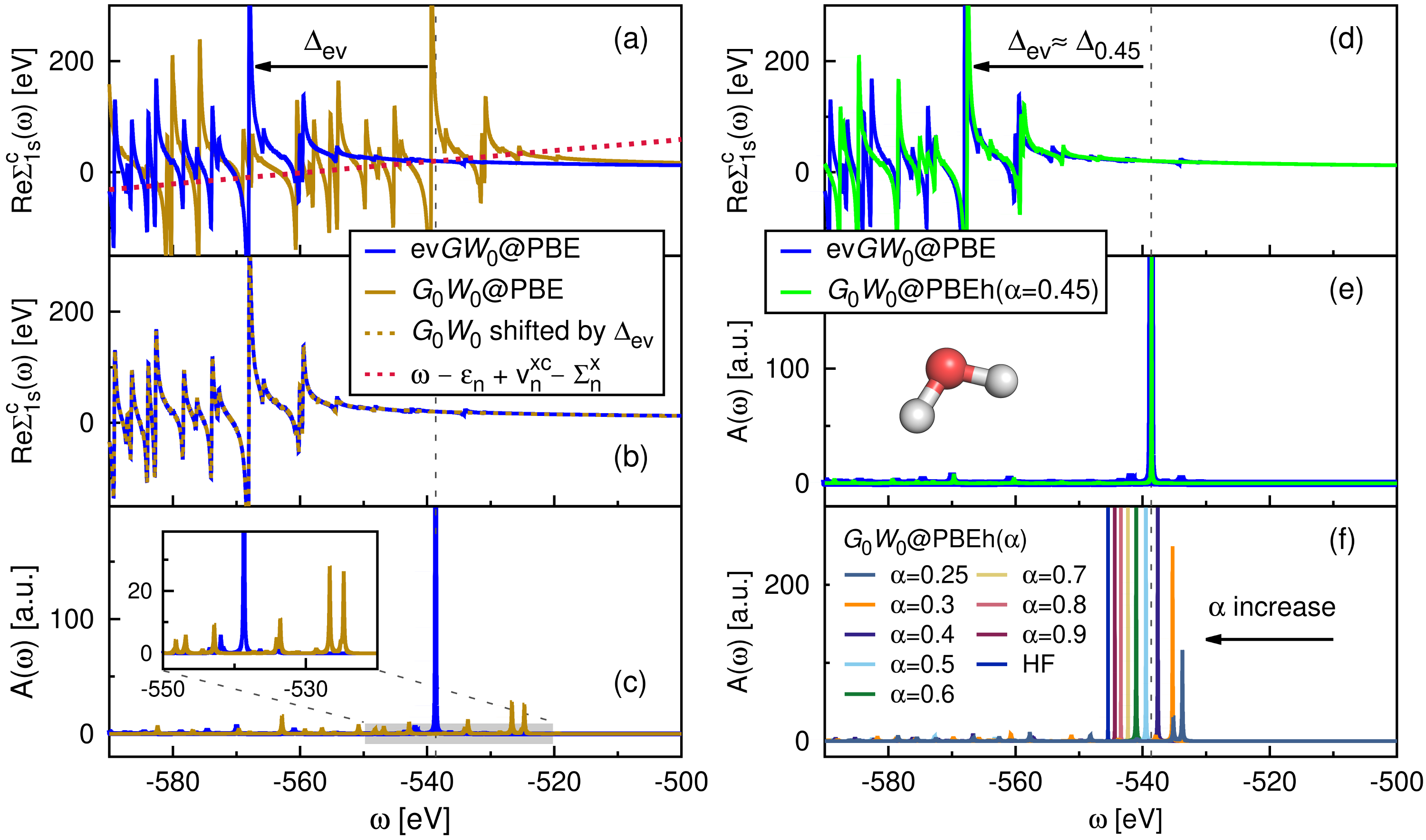}
  \caption{Core excitation for a single water molecule from $G_0W_0$ and ev$GW_0$. (a) Real part of the self-energy $\Sigma^c(\omega)$ (correlation contribution) 
  using the PBE functional as starting point. Diagonal matrix elements $\Re\Sigma^c_n(\omega)=\braket{\psi_n|\Re\Sigma^c(\omega)|\psi_n}$ for the oxygen 1s orbital. (b) Self-energy from 
  $G_0W_0@$PBE shifted by $\Delta_{\textnormal{ev}}$ relative to the ev$GW_0$ result.  The intersection with the red dashed line is the graphical solution of
  the QP Equation \eqref{eq:qpe}. (c) Spectral function $A(\omega)$ from $G_0W_0@$PBE and ev$G_0W_0$@PBE. (d) Self-energy and (e) spectral function using PBEh($\alpha=0.45$) 
  as starting point. (f) Spectral function obtained from $G_0W_0@$PBEh for different amounts of exact exchange $\alpha$. The vertical gray-dashed line indicates the QP solution from 
ev$GW_0@$PBE. Note that the self-energy is slightly broadened for better visualization and that each sharp peak actually corresponds to a pole.}
   \label{fgr:selfenergy_spectral}
\end{figure*}
The $GW$ approximation to many-body perturbation theory \cite{Hedin1965} is a promising method to improve upon the limitations of traditional $\Delta$-approaches and has become a widespread tool for the accurate prediction of electron removal energies of valence states in molecular and solid-state systems.\cite{Golze2019} $GW$ is routinely applied to systems 
with several hundred atoms,\cite{Govoni2015,Wilhelm2016,Wilhelm2017} and recently even to system sizes with more than 1000 atoms.\cite{Wilhelm2018,DelBen2019} However, core-level spectroscopy has been 
rarely attempted with $GW$. Recently, the first promising results were obtained for solid-state systems.\cite{Zhou2015,Aoki2018} The few existing studies for molecular core excitations give a mixed 
first impressions since anything between 0.5~eV\cite{Golze2018} and 10~eV deviation from experiment has been reported.\cite{Setten2018,Voora2019} In this work, we show how reliable and highly accurate 
core-level BEs can be obtained from $GW$ and explain why large deviations from experiment were reported earlier. We also present a $GW$ benchmark set for 1s core states complementary to the 
popular $GW$100 benchmark set\cite{Setten2015} for valence excitations. \par
First, we introduce the $GW$ framework.
The central object of $GW$ is the self-energy $\Sigma$, which contains all quantum mechanical exchange and correlation interactions of the hole created by the excitation process 
and its surrounding electrons. The self-energy is calculated from the Green's function $G$ and the perturbation expansion in the screened Coulomb interaction $W$ as formulated by 
Hedin in the 1960s~\cite{Hedin1965}. The poles of $G$ directly correspond to the excitation energies as measured in photoemission spectroscopy. 

In practice, $GW$ is performed within the first-order perturbation theory ($G_0W_0$) and starts from a set of mean-field single-particle orbitals $\{\psi_n\}$ and corresponding eigenvalues $\{\varepsilon_n\}$. 
These are usually obtained from a preceding KS-DFT or Hartree-Fock (HF) calculation. 
The $GW$ quasiparticle (QP) energies $\varepsilon_n^{G_0W_0}$ are computed by iteratively solving 
\begin{equation}
\label{eq:qpe}
  \varepsilon_{n}^{G_0W_0}=\varepsilon_{n}+\mathrm{Re}\Braket{\psi_{n}|\Sigma\left(\varepsilon_{n}^{G_0W_0}\right)-v^{\rm xc}|\psi_{n}},
\end{equation}
for $\varepsilon_{n}^{G_0W_0}$, where $v^{\rm xc}$ is the XC potential from DFT and spin variables are omitted. In the following, we use the notation $\Sigma_n=\Braket{\psi_n|\Sigma|\psi_n}$ and 
$v^{xc}_n=\Braket{\psi_n|v^{xc}|\psi_n}$ for the ($n,n$) diagonal matrix elements of the self-energy and XC potential. The QP energies are related to the BE of state $n$ by 
$\text{BE}_n=-\varepsilon_n^{G_0W_0}$ and the self-energy $\Sigma$ is given by
\begin{equation}
\label{eq:self_energy}
 \Sigma(\mathbf{r},\mathbf{r}',\omega)=
\frac{i}{2\pi}\int  d\omega'  e^{i\omega'\eta} 
G_0(\mathbf{r},\mathbf{r}',\omega+\omega')W_0(\mathbf{r},\mathbf{r}',\omega')
\end{equation}
where $\eta$ is a positive infinitesimal.
The self-energy is typically split into a correlation $\Sigma^c$ and an exchange part 
$\Sigma^x$, $\Sigma=\Sigma^c+\Sigma^x$, where $\Sigma^c$ is computed from $W_0^c=W_0-v$ and $\Sigma^x$ from the bare 
Coulomb interaction $v$.
The mean-field Green's function $G_0$ is given by
\begin{equation}
 G_0(\mathbf{r},\mathbf{r}',\omega) = \sum_m\frac{\psi_{m}(\mathbf{r})\psi_{m}(\mathbf{r}')}{\omega-\varepsilon_{m}-i\eta 
\sgn(\varepsilon_{\rm F}-\varepsilon_{m})},
 \label{eq:greensfkt}
\end{equation}
 where $\varepsilon_{\rm F}$ denotes the Fermi energy. $W_0$ in Equation~\eqref{eq:self_energy} is the screened Coulomb 
interaction in the random-phase approximation (RPA) and is computed from the dielectric 
function as described in Ref.~\citenum{Golze2019}.\par
We now discuss the application of $GW$ to core-level spectroscopy. The basic requirement to obtain computational XPS data from $GW$ is an explicit description of the core electrons. We treat the latter efficiently by working in a local 
all-electron basis of numeric-atomic centered orbitals (NAOs). Furthermore, we showed that highly accurate frequency integration techniques for the computation of the 
self-energy (Equation~\eqref{eq:self_energy}) are required for core states.\cite{Golze2018} Unlike for valence states, the self-energy has a complicated structure with many poles 
in the core region, as displayed in Figure~\ref{fgr:selfenergy_spectral}(a). For such complex pole structures, the analytic continuation, that is frequently employed in $GW$ calculations for valence states to continue $\Sigma^c$ from the imaginary to the real frequency axis, fails completely.\cite{Golze2018} We showed 
that the contour deformation (CD) technique, in which a full-frequency integration on the real frequencies axis is performed, yields the required accuracy. Results from CD exactly 
match the computationally demanding fully analytic solution of Equation~\eqref{eq:self_energy}.\cite{Golze2018} Our CD-$GW$ implementation is computationally efficient enabling 
the computation of system sizes exceeding 100 atoms, see Ref.~\citenum{Golze2018} for details of our $GW$ core-level implementation in the all-electron code 
FHI-aims.\cite{Blum2009}
Numerically stable and precise algorithms for the computation of the self-energy are only the first step toward reliable core-level excitations from 
$GW$. In the following, the failure of standard $G_0W_0$ schemes for core states is explored. \par
Figure~\ref{fgr:selfenergy_spectral}(a) shows the $G_0W_0$ self-energy matrix elements for the O1s state of an isolated water molecule using the Perdew-Burke-Ernzerhof 
(PBE)\cite{Perdew1996} functional for the underlying DFT calculation ($G_0W_0@$PBE). Instead of iterating Equation~\eqref{eq:qpe}, we can obtain its solution graphically by 
finding the intersections of the straight line $\omega-\varepsilon_n+v_n^{xc}-\Sigma_n^x$ with the self-energy matrix elements $\Sigma_{\rm n}^c$. As apparent from 
Figure~\ref{fgr:selfenergy_spectral}(a), a clear single solution is missing. Many intersections are observed, which are all valid solutions of Equation~\eqref{eq:qpe}. 

To further investigate this multi-solution behavior, we calculate the spectral function $A(\omega)$\cite{Golze2018,Golze2019}
\begin{equation}
A(\omega)=\frac{1}{\pi}\sum_m\frac{\left|\Im\Sigma_m(\omega)\right|}{\left[\omega-\varepsilon_m-\left(\Re\Sigma_m(\omega)-v_m^{xc}\right)\right]^2+\left[\Im\Sigma_m(\omega)\right]^2}
 \label{eq:Aomegatrace}
\end{equation}
where we include also the imaginary part of the complex self-energy and use, unlike in fully self-consistent $GW$,\cite{Caruso2012,Caruso2013} only the diagonal matrix elements of $\Sigma$. The spectral function for the oxygen 1s excitation of an isolated water molecule is reported in Figure~\ref{fgr:selfenergy_spectral}(c). We observe many peaks with similar spectral weight. No distinct peak can be assigned to the QP excitation. In other words, $G_0W_0@$PBE does not provide a unique QP solution for the 1s excitation. This is in sharp contrast to the valence case, where 
$G_0W_0@$PBE is routinely applied to molecules and a clear single solution has been reported in the vast majority of cases.\cite{Setten2015}. 

Figure~\ref{fgr:selfenergy_spectral}(c) illustrates that for core states, the QP energy and the satellite spectrum have merged. Satellites are, e.g., due to multi-electron excitations such as shake-up processes\cite{Sankari2006,Schirmer1987} and have typically much smaller spectral weights than the QP peak. The fact that we observe the opposite for $G_0W_0@$PBE implies that almost all spectral weight has been transferred from the QP peak to the satellites. We will next investigate the origin of this behavior and provide a solution. \par
We start by updating the KS eigenvalues $\{\varepsilon_m\}$ in the Green's function with the $G_0W_0$ quasiparticle energies, re-evaluate Equation~\eqref{eq:qpe} and iterate until $G$ is self-consistent in the eigenvalues. For most valence and virtual states, a unique QP solution exists at the $G_0W_0$ level, while for core states we initialize the iteration in $G$ with an approximation of the QP energy. This procedure yields a partially eigenvalue self-consistent scheme denoted as ev$GW_0@$PBE, where $W$ is kept fixed at the $W_0$ level. Iterating the eigenvalues in $G$ shifts the onset of the pole structure of the self-energy to lower energies, see Figure~\ref{fgr:selfenergy_spectral}(a). The pole structure of the self-energy looks similar in $G_0W_0@$PBE and ev$GW_0@$PBE, but shifted by a constant amount. Figure~\ref{fgr:selfenergy_spectral}(b) shows, that the $G_0W_0@$PBE self-energy is indeed almost identically to ev$GW_0@$PBE when shifted by $\Delta_{\rm ev}=-28.7$~eV.  The effect of this shift is that the graphical solution now produces a clear QP solution and a satellite spectrum with much lower intensity, as displayed in Figure~\ref{fgr:selfenergy_spectral}(c). In other words, eigenvalue self-consistency in $G$ achieves a separation of QP peak and satellite spectrum for deep core states. This eigenvalue self-consistency strategy was already employed for $3d$ states in transition metal oxides\cite{Gatti2015,Byun2019} or semi-core states in sodium\cite{Zhou2015} and can be understood as follows.\par

Satellites occur in frequency regions, where the real part of $\Sigma^c_n$ has poles and its imaginary part complementary peaks, which is shown in detail in our recent $GW$ review article.\cite{Golze2019} As obvious from Equation~\eqref{eq:Aomegatrace}, large imaginary parts correlate with low spectral weights, i.e., satellite character. Rewriting the self-energy into analytic form reveals its pole-structure
\begin{equation}
 \Sigma_n^c(\omega) = \sum_m \sum_{s} \frac{\Braket{\psi_n\psi_m|P_s|\psi_m\psi_n}}{\omega -\varepsilon_m + 
(\Omega_s-i\eta)\sgn(\varepsilon_{\mathrm{F}} -\varepsilon_m)},
    \label{eq:sigma_pole}
\end{equation}
where $\Omega_s$ are charge neutral excitations and $P_s$ transition amplitudes\cite{Golze2019}. The $G_0W_0$ self-energy therefore has poles at $\varepsilon_{i} -\Omega_{s}$ and $\varepsilon_{a} +\Omega_{s}$, where $i$ indicates occupied and $a$ virtual states. Each of these poles gives rise to satellite features and can be understood as an electron or hole excitation coupled to a neutral excitation.\par
For $G_0W_0$@PBE, $\varepsilon_{i}$ are PBE eigenvalues and $\Omega_s$ are close to PBE eigenvalue differences between occupied and virtual states. The neutral excitations $\Omega_s$ are typically underestimated at the PBE level, while the eigenvalues are overestimated by several eV in the valence region of the spectrum and by 20 to 30~eV for the 1s core states.\cite{Golze2018,Setten2018} The $\varepsilon_{\rm 1s}-\Omega_{s}$ poles in the self-energy are therefore considerably too high in energy and start to energetically overlap with the QP energy of the core state. This explains why the satellites have such high spectral weight in $G_0W_0$@PBE and why no distinct QP peak can be found.\par
 In ev$GW_0$, we replace the KS-DFT eigenvalues $\{\varepsilon_m\}$ in Equation~\eqref{eq:greensfkt} by $\varepsilon_{m}+\Delta \varepsilon_{m}$,  where $\Delta \varepsilon_{m}$ is the $GW$ correction. For a PBE starting point, $\Delta\varepsilon_m$ is negative for occupied states and the poles of $\Sigma^c_n$ shift to lower energies, away from the QP energy. The poles in the core region are now located at $\varepsilon_{1s}+\Delta\varepsilon_{1s}-\Omega_s$ and the corresponding satellite peaks are separated from the QP peak and reduced in spectral weight.\par
The effect of self-consistency in $G$ can be reproduced in a $G_0W_0$ calculation, which is computationally less demanding, by including exact exchange in the DFT functional. 
We employ the PBE-based hybrid (PBEh) functional family \cite{Atalla2013}, which is characterized by an adjustable fraction $\alpha$ of HF exchange and 
corresponds for $\alpha=0.25$ to the PBE0\cite{Adamo1999,Ernzerhof1999} functional. For $\alpha=0.45$, we obtain approximately the same shift of the pole structure as in 
ev$GW_0@$PBE and observe a distinct QP peak at the same frequency, see Figure~\ref{fgr:selfenergy_spectral}(d) and 
(e). Increasing the amount of exact exchange, the QP peak in the spectral function moves to lower energies, which is in agreement with the starting point optimization 
studies conducted for
valence excitations.\cite{Marom2012,Atalla2013,Golze2019} However, distinct QP peaks are only obtained for $\alpha > 0.3$. As shown Figure~\ref{fgr:selfenergy_spectral}(f), $G_0W_0@$PBE0 still suffers from a large transfer of spectral weight to the satellites.\par
Previous $GW$ core-level studies for small molecules \cite{Setten2018,Voora2019} reported $G_0W_0$ calculations performed on top of generalized gradient approximation (GGA) or hybrid functionals with a low amount of exact exchange. Our analysis presented in this article demonstrates that those studies cannot have found the QP solution because their spectral function would look like the yellow spectrum in Figure~\ref{fgr:selfenergy_spectral}(c). Linearizing the QP equation by a Taylor expansion to first-order around $\varepsilon_n$, as done in  Refs.~\citenum{Voora2019}, \citenum{Setten2018} and \citenum{Aoki2019}, for such a spectral function leads to uncontrollable results, which partly explains the large deviation of the reported results from experiment\cite{Setten2018,Voora2019}. Furthermore, the linearization error increases rapidly with increasing binding energy and may already amount to 0.5~eV for deeper valence states, as shown in Ref.~\citenum{Golze2019}. As already pointed out in our previous work\cite{Golze2018}, 
Equation~\eqref{eq:qpe} should always be solved iteratively for core states. \par
\begin{figure}[t]
\centering
  \includegraphics[width=0.99\columnwidth]{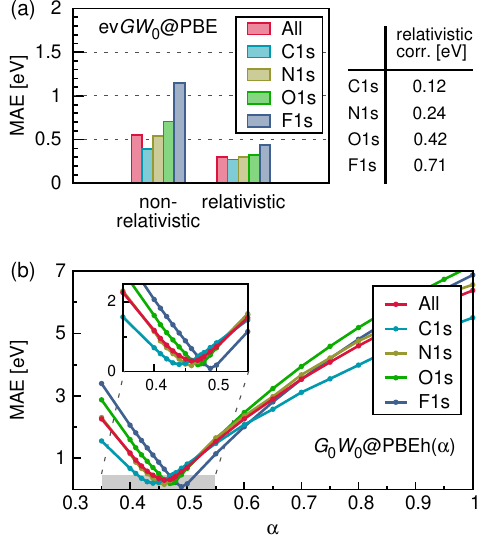}
  \caption{Mean absolute error (MAE) of the absolute BEs for the CORE65 benchmark set with respect to experiment. (a) ev$GW_0@$PBE with and without relativistic correction. (b) MAE for $G_0W_0@$PBEh
dependent on the amount of exact exchange $\alpha$ in the PBEh functional. Relativistic corrections are included.}
 \label{fgr:evgw0_rel}
\end{figure}
We now assess the accuracy of ev$GW_0@$PBE and $G_0W_0@$PBEh with respect to experiment for a benchmark set of 65 1s binding energies of gas-phase molecules, denoted in the following as CORE65. This benchmark set contains 30 C1s, 21 O1s, 11 N1s and 3 F1s excitations from 32 small, inorganic and organic molecules up to 14 atoms, see Table~S1 in the Supporting Information (SI) for details. The CORE65 benchmark covers a variety of different chemical environments and bonding types and the most common functional groups. As with all correlated electronic structure methods, $GW$ converges slowly with respect to basis set size.\cite{Golze2019} Even at the quadruple-$\zeta$ level, the BEs deviate by 0.2~eV to 0.4~eV from the complete basis set limit (see Tables~S2 and S3 in SI). All $GW$ 
results are thus extrapolated to the complete basis set limit using the Dunning basis set family cc-pV$n$Z ($n$=3-6).\cite{Dunning1989,Wilson1996}  \par
Since we expect relativistic effects to become important for heavier elements, we add relativistic corrections for the 1s excitations as post-processing step to the $GW$ calculation. Our relativistic corrections have been obtained by solving the radial KS and 4-component Dirac-KS equations self-consistently for a free neutral atom at the PBE level, and evaluating the difference between their 1s eigenvalues; see Figure~\ref{fgr:evgw0_rel}(a).  Details of our relativistic correction scheme, which is similar to the one reported in Ref.~\citenum{Bellafont2016b}, and its comparison to other relativistic methods will be described in a forthcoming paper.\par
For ev$GW_0@$PBE, the mean absolute error (MAE) of the absolute BEs with respect to experiment is reported in Figure~\ref{fgr:evgw0_rel}(a) for relativistic and non-relativistic calculations. We find that relativistic effects start to dominate the error in the QP energies already for second-row elements. In the non-relativistic case, the core-level BEs are generally underestimated (see Table S2 in the SI) and the MAE increases with the atomic number. Accounting for relativistic effects, the species dependence in the MAE is largely eliminated. \par

Figure~\ref{fgr:evgw0_rel}(b) shows the MAE at the $G_0W_0@$PBEh($\alpha$) level with respect to the amount of exact exchange $\alpha$ in the PBEh functional, including relativistic corrections. These $\alpha$ dependent calculations are performed for a subset of 43 excitations of the CORE65 benchmark set, for which the mapping between core state and atom is trivial and requires no analysis of, e.g., molecular orbital coefficients. The smallest MAE is obtained for $\alpha$ values around 0.45. This observation agrees nicely with our analysis of the self-energy in Figure~\ref{fgr:selfenergy_spectral}(d), where we found that $\alpha\approx0.45$ reproduces the ev$GW_0$ self-energy best. For smaller $\alpha$ values, the BEs are underestimated and for larger values increasingly overestimated. The species dependence of the optimal $\alpha$ values are mostly reduced when taking relativistic effects into account. The optimal $\alpha$ values increase only slightly with the atomic number ranging from 0.44 to 0.49, see 
Figure~\ref{fgr:evgw0_rel}(b). \par
\begin{figure}[t]
\centering
  \includegraphics[width=0.99\columnwidth]{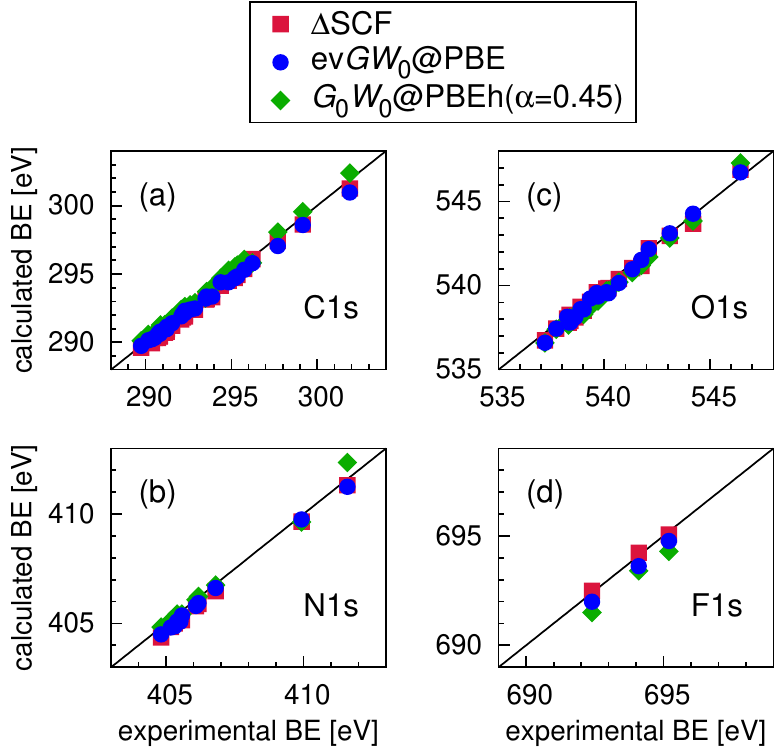}
  \caption{Absolute C1s (a), N1s (b), O1s (c) and F1s (d) binding energies (BEs) for the CORE65 benchmark set comparing calculated values at the $\Delta$SCF, ev$GW_0@$PBE and 
$G_0W_0@$PBEh($\alpha=0.45$) level to experiment. The respective computational method underestimates the BE when the data point is below the black line and overestimates when   
 above. }
 \label{fgr:abs_vs_exp}
\end{figure}
\begin{table*}[t]
  \fontsize{10}{12}\selectfont
  \center
  \caption{Mean absolute error (MAE) in [eV] with respect to experiment for the CORE65 benchmark set. MAE for absolute and relative core-level BEs, where the latter is the shift of the BE 
with respect to a reference molecule. \ce{CH4}, \ce{NH3}, \ce{H2O} and \ce{CH3F} have been used as reference molecules for C1s, N1s, O1s and F1s respectively. Relativistic effects are accounted for 
in all three 
methods. }
  \begin{tabular*}{0.99\linewidth}{@{\extracolsep{\fill}}ccccccccccc}\toprule
  & \multicolumn{2}{c}{$\Delta$SCF}  & \multicolumn{2}{c}{ev$GW_0@$PBE}  & \multicolumn{2}{c}{$G_0W_0@$PBEh($\alpha$=0.45)} \\\cmidrule(l{0.5em}r{0.5em}){2-3}  
\cmidrule(l{0.5em}r{0.5em}){4-5}\cmidrule(l{0.5em}r{0.5em}){6-7}
\centering{core-level}  & absolute BEs & relative BEs  & absolute BEs & relative BEs  & absolute BEs & relative BEs  
\\\hline

all & 0.33 & 0.14 & 0.30 & 0.18 & 0.33  &  0.26  \\
C1s & 0.36 & 0.10 & 0.27 & 0.18 & 0.24  &  0.29 \\
N1s & 0.32 & 0.08 & 0.30 & 0.14 & 0.16  &  0.23 \\
O1s & 0.32 & 0.22 & 0.32 & 0.22 & 0.48  &  0.25 \\
F1s & 0.12 & 0.13 & 0.44 & 0.05 & 0.83  &  0.11  
\\\bottomrule
  \end{tabular*}
  \label{tab:mae}
\end{table*}
Figure~\ref{fgr:abs_vs_exp} compares the absolute BEs obtained from experiment to the theoretical BEs computed at the $\Delta$SCF, ev$GW_0@$PBE and $G_0W_0@$PBEh($\alpha=0.45$) level. The $\Delta$SCF 
calculations are performed with the PBE0\cite{Adamo1999,Ernzerhof1999} functional and are carefully converged adding 
additional tight basis functions to standard Gaussian basis sets for the core-hole 
calculation.\cite{Ambroise2019} Following a recently proposed $\Delta$SCF simulation protocol\cite{Kahk2019}, we include scalar relativistic effects self-consistently via the zeroth order 
regular approximation (ZORA).\cite{Lenthe1994} Our BEs obtained from $\Delta$SCF-PBE0 agree with an overall MAE of 0.33~eV  much better with experiment than reported in previous 
studies (0.7~eV)\cite{Bellafont2015}, which must be attributed to incomplete basis sets and the neglect of relativistic effects. \par
The ev$GW_0@$PBE approach yields excellent agreement of the absolute BEs with experiment consistently for all data points as shown in Figure~\ref{fgr:abs_vs_exp}. With an overall MAE of 0.3~eV, the 
accuracy of ev$GW_0@$PBE is well within the chemical resolution required for the interpretation of most XPS spectra. $G_0W_0@$PBEh($\alpha=0.45$) yields a similar overall MAE, which, 
however, depends to some extent on the species, see Table~\ref{tab:mae}. As shown in Figure~\ref{fgr:abs_vs_exp}(d), F1s removal energies are systematically underestimated with $G_0W_0@$PBEh. Results for this element could in principle be improved by using an element-specific optimized $\alpha$ value, based on the analysis in Figure~\ref{fgr:evgw0_rel}(b).\par
Relative BEs are very well reproduced with all three theoretical methods, as shown in Table~\ref{tab:mae} and in more detail in Figure~S1 (SI). With $\Delta$SCF and 
ev$GW_0@$PBE we obtain MAEs smaller than 0.2~eV and slightly larger errors between 0.2 and 0.3~eV with $G_0W_0@$PBEh($\alpha=0.45$). Results for F1s are 
reported for the sake of completeness. However, note that we have only two data points for the relative BEs and the experimental uncertainties are generally  
larger for fluorine than for the lighter elements. Except for F1s BEs, the MAEs are not species dependent.\par
In summary, we showed that $GW$ is a reliable and accurate method to calculate 1s core excitations. However, standard 
$G_0W_0$ setups routinely used for valence excitations cannot be employed. For core states, $G_0W_0$ calculations starting from GGA or standard hybrid functionals experience a huge 
weight transfer from the quasiparticle to the satellites. In fact, this weight transfer is so extreme that a unique QP solution does not exist for the molecules we have investigated. We demonstrated for a PBE starting point
that eigenvalue self-consistency in $G$ is mandatory to achieve a proper separation between QP and satellite peaks in the $GW$ calculation. The effects of ev$GW_0$ can be reproduced in 
$G_0W_0$, which is computationally less expensive, by using a hybrid functional with a high fraction of exact exchange as starting point. We found that 45\% of HF exchange is 
optimal. Furthermore, the inclusion of relativistic effects and a proper extrapolation to the complete basis limit are crucial to obtain accurate core-level BEs.  Our work is an important stepping stone for the accurate calculation of XPS spectra of condensed systems, where $\Delta$-based 
approaches face conceptual limitations. $GW$ can be applied without restrictions to systems with periodic boundary conditions and is also for large molecular structures a reliable 
and numerically robust method. Furthermore, this work is fundamental for the calculation of X-ray absorption spectra (XAS) from the Bethe-Salpeter 
equation\cite{Salpeter1951}, which uses the $GW$ results as input.

\section*{Computational Details}
\label{sec:computational}
All calculations are performed with the FHI-aims program package~\cite{Blum2009,Havu2009,Ren2012}, where the all-electron KS equations are solved in the NAO scheme. The structures 
of the CORE65 molecules have been optimized at the DFT level using NAOs of \textit{tier 2} quality~\cite{Blum2009} to represent core and valence electrons. The PBE functional\cite{Perdew1996} is used 
to model exchange and correlation in combination with the atomic ZORA~\cite{Lenthe1994,Blum2009} kinetic energy operator. Van der Waals interactions are accounted for by 
employing the Tkatchenko-Scheffler dispersion correction.\cite{Tkatchenko2009}\par
Core-level BEs from $\Delta$SCF are calculated using the PBE0\cite{Adamo1999,Ernzerhof1999} hybrid functional, (atomic) ZORA and 
def2 quadruple-$\zeta$ valence plus polarization (def2-QZVP)\cite{Weigend2005} basis sets. The def2-QZVP basis sets are all-electron basis sets of contracted Gaussian orbitals, which are optimized 
to yield accurate total energies\cite{Weigend2005}. Gaussian basis sets can be considered as a special case of an NAO and are treated numerically in FHI-aims. To guarantee the 
full relaxation of other electrons in the presence of the core hole, we decontracted the def2-QZVP basis sets in the $\Delta$SCF calculation to add tighter core functions. To 
properly localize the core hole at a specific atom, we performed a Boys localization\cite{Foster1960} at the end of the SCF cycle of the charge neutral calculation and used this 
wavefunction as initial guess for the charged system. \par
For the $GW$ calculations, we use the contour deformation technique\cite{Govoni2015,Golze2018,Golze2019,Holzer2019} to 
evaluate the frequency integral of the self-energy and employ a modified Gauss-Legendre grid\cite{Ren2012} with 200 grid points for the imaginary frequency integral. The QP 
equation is \textit{always} solved iteratively. For ev$GW_0$, we iterate additionally  the QP energies in $G$ including explicitly all occupied states and the first 
five virtual states in the iteration. Scissor shifts are employed for the remaining virtual states. For the partially self-consistent ev$GW_0$ calculations, we use the PBE functional as starting point and for $G_0W_0$@PBEh($\alpha$) calculations the PBEh($\alpha$) hybrid functionals\cite{Atalla2013}. The core-level BEs are extrapolated to the complete basis set limit using the Dunning basis set family cc-pV$n$Z ($n$=3-6) \cite{Dunning1989,Wilson1996}, which are standard basis sets for correlated electronic-structure methods. The extrapolation has been performed with four points by a linear regression against the 
inverse of the total number of basis functions. The standard error of the extrapolation is smaller than 0.1~eV and the correlation coefficient $R^2$ in most cases $>$ 0.9, see 
Table~S2 and S3 in the SI. Alternatively, the extrapolation can be performed with respect to $C_n^{-3}$, where $C_n$ is the cardinal number of the basis set. We found that the difference between both extrapolation schemes is very small, e.g., the average absolute deviation is only 0.04~eV for the CORE65 $G_0W_0@$PBEh($\alpha$=0.45) data. Self-energy matrix elements and spectral functions are calculated at the cc-pV4Z level. The relativistic corrections for the $GW$ energies are computed at the PBE level from free neutral atom calculations on 
numerical real space grids.\cite{Certik2013}

%
\begin{acknowledgement}
\fontsize{10}{12}\selectfont
 We thank the CSC - IT Center for Science for providing computational resources. D. Golze acknowledges financial support by the Academy of Finland (Grant 
No. 316168). We thank Lucia Reining, Michiel van Setten and Volker Blum for fruitful discussions.
\end{acknowledgement}
 
\begin{suppinfo}
\fontsize{10}{12}\selectfont
Plot of relative core-level BEs comparing theory and experiment (Figure~S1); CORE65 benchmark results at the $\Delta$SCF, ev$GW_0@$PBE and $G_0W_0@$PBEh($\alpha$=0.45) level including relativistic 
corrections and experimental data (Table~S1 and Figure~S2);  non-relativistic ev$GW_0@$PBE and $G_0W_0@$PBEh($\alpha$=0.45) results for basis set series cc-pV$n$Z ($n$=3-6), 
extrapolated values, standard errors and correlation coefficients (Table~S2 and S3); structures of CORE65 benchmark set. 

\end{suppinfo}


\bibliography{ref}

\end{document}